\documentclass[conference]{IEEEtran}
\IEEEoverridecommandlockouts
\usepackage{amsmath,amssymb,amsfonts}
\usepackage{mathptmx}
\usepackage{algorithmic}
\usepackage{graphicx}
\usepackage[numbers,sort&compress]{natbib}
\usepackage{hyperref}
\begin{document}

\title{An Opportunistic-Non Orthogonal Multiple Access based Cooperative Relaying System over Rician Fading Channels}

\author{\IEEEauthorblockN{Pranav Kumar Jha$^{\href{orcid.org/0000-0001-8053-988X}{*}}$, {S Sushmitha Shree$^{\href{orcid.org/0000-0003-4929-7776}{\#}}$}, \textit{IEEE Member} and {D. Sriram Kumar$^{}$}}
\IEEEauthorblockA{\textit{$^{}$Department  of ECE, {National Institute of Technology, Tiruchirappalli, Tamil Nadu, India}} \\
\textit{$^{\#}$Department of ECE, {Thiagarajar College of Engineering, Madurai, Tamil Nadu, India}} \\
$^{*}$\textit{E-Mail} : jha\_k.pranav@live.com}
}
\maketitle

\begin{abstract}
Non-orthogonal Multiple Access (NOMA) has become a salient technology for improving the spectral efficiency of the next generation 5G wireless communication networks. In this paper, the achievable average rate of an Opportunistic Non-Orthogonal Multiple Access (O-NOMA) based Cooperative Relaying System (CRS) is studied under Rician fading channels with Channel State Information (CSI) available at the source terminal. Based on CSI, for opportunistic transmission, the source immediately chooses either the direct transmission or the cooperative NOMA transmission using the relay, which can provide better achievable average rate performance than the existing Conventional-NOMA (C-NOMA) based CRS with no CSI at the source node. Furthermore, a mathematical expression is also derived for the achievable average rate and the results are compared with  C-NOMA based CRS with no CSI at the transmitter end, over a range of increasing power allocation coefficients, transmit Signal-to-Noise Ratios (SNRs) and average channel powers. Numerical results show that the CRS using O-NOMA with CSI achieves better spectral efficiency in terms of the achievable average rate than the Conventional-NOMA based CRS without CSI. To check the consistency of the derived analytical results, Monte Carlo simulations are performed which verify that the results are consistent and matched well with the simulation results. 
\end{abstract}

\begin{IEEEkeywords}
Non-orthogonal Multiple Access, Achievable Rate, Channel State Information, Rician Fading Channels, Decode-and-Forward Relay, Signal-to-Noise Ratio
\end{IEEEkeywords}

\IEEEpeerreviewmaketitle

\section{Introduction}
\label{sec1}
Non-orthogonal multiple access (NOMA) is one of the significant techniques to enhance the spectral efficiency of the wireless multi-user communication systems for the fifth generation (5G) networks \citep{ding2014performance,dai2015non,fang2016energy,sun2017optimal,ding2016relay}. In industry, the NOMA and the cooperative relaying scheme have been considered as primary techniques for 3rd Generation Partnership Project (3GPP) Long Term Evolution-Advanced (LTE-A) systems \citep{do2009linear,bai2013evolved}. Distinct from conventional orthogonal-multiple access (OMA) techniques, NOMA enables various users for simultaneous transmission of signals with the same channel resources in time or frequency domain but with different levels of power \citep{dai2015non,fang2016energy,ding2016relay,sun2017optimal}. 
\subsection{Basic Principle of NOMA and Related works}
The key principle of NOMA is to utilize the additional or unused power domain to further in the number of users where users with good channel conditions are used as relays to improve the performance of the system with the help of successive interference cancellation (SIC). Characteristically, users are classified according to their channel conditions as strong users and weak users, where strong users have good channel conditions and weak users have bad channel conditions. It is obvious to allocate less power to
strong users at the source end. In this fashion, the superposition of signals are sent from the transmitter end with distinct power levels
and the receiver incorporates SIC for the realization of multi-user detection \citep{saito2013system,liu2016capacity,sun2015ergodic}, which allows NOMA to
entertain more users making it one of the promising techniques for massive connectivity in the 5G networks but at the cost of
tractable increased complexity in the design of receiver due to the presence of 
SIC in NOMA systems \citep{saito2013system}. In order to enable users to serve as cooperative relays, they must be priorly notified of the messages of other users which can be achieved by the use of SIC.
Over and above, the
system performance of cellular networks can be enhanced significantly by using cooperative relaying itself. So, the combination of the cooperative relaying system with NOMA is highly encouraging for the enhancement of the
throughput for the upcoming 5G wireless networks, which has captivated growing interests, recently. In particular, a cooperative relay transmission using NOMA was suggested in \citep{ding2015cooperative}, where
strong users decode the signals that are destined to other users and be employed as relays to boost the performance of weak users.
Another cooperative relaying scheme using NOMA was proposed in
\citep{kim2015capacity}, where the performance analysis with decode-and-forward (DF) relay over Rayleigh fading channels was
investigated. 
 
\subsection{Motivation and Contribution}
Recently, in \cite{lee2017achievable}, an Opportunistic-NOMA (O-NOMA) based
CRS is proposed using half-duplex DF relay under Rayleigh fading channels and the achievable average rate has been studied for different channel powers and power allocation coefficients with channel state information (CSI) known at the source end. In general, most of the techniques were analyzed in
systems only under Rayleigh fading channels to improve the spectral efficiency \citep{kim2015capacity} and very few have considered to use Rician fading channels which can fit better with some of the typical 5G applications. This is the motivation towards this work which adds extra contribution in cooperative relaying schemes based on NOMA.

In this paper, the Opportunistic-NOMA (O-NOMA) based CRS system model \cite{lee2017achievable} is considered, but Rayleigh fading channels is replaced with Rician fading channels to analyze spatially multiplexed transmissions, which provides novelty to this work in determining better spectral efficiency for cooperative O-NOMA. It is also assumed that CSI is available at the source end in order to achieve further improvement in the performance but, at the expense of system overhead and complexity, where a source, a half-duplex decode-and-forward relay, and a destination are considered for the implementation purpose. For opportunistic transmission, CSI for the source-to-relay and source-to-destination links are exploited and transmitter instantaneously decides the best transmission path from the direct transmission and the cooperative NOMA transmission using relays. In addition, a general mathematical expression of the achievable average rate is also deduced and the results are verified through Monte Carlo simulations. The simple notations of O-NOMA and C-NOMA have been used to indicate the cooperative relaying system based on Opportunistic-NOMA with CSI and Conventional-NOMA without CSI, respectively.
\subsection{Structure}
The organization of the work is as follows: Section \ref{sec2} provides system description of O-NOMA and also provides
its received signals and signal-to-noise ratios (SNRs). In Section {\ref{sec3}}, the achievable average rate of O-NOMA is examined and an expression is derived for the total achievable average rate. In Section \ref{sec4}, the achievable average rate of the O-NOMA is compared with the C-NOMA presented in \citep{7983401} for different power allocation coefficients and average channel powers used for NOMA through numerical results and simulations. Conclusions are drawn in Section \ref{sec5}.  

\begin{figure}
\includegraphics[width=\linewidth]{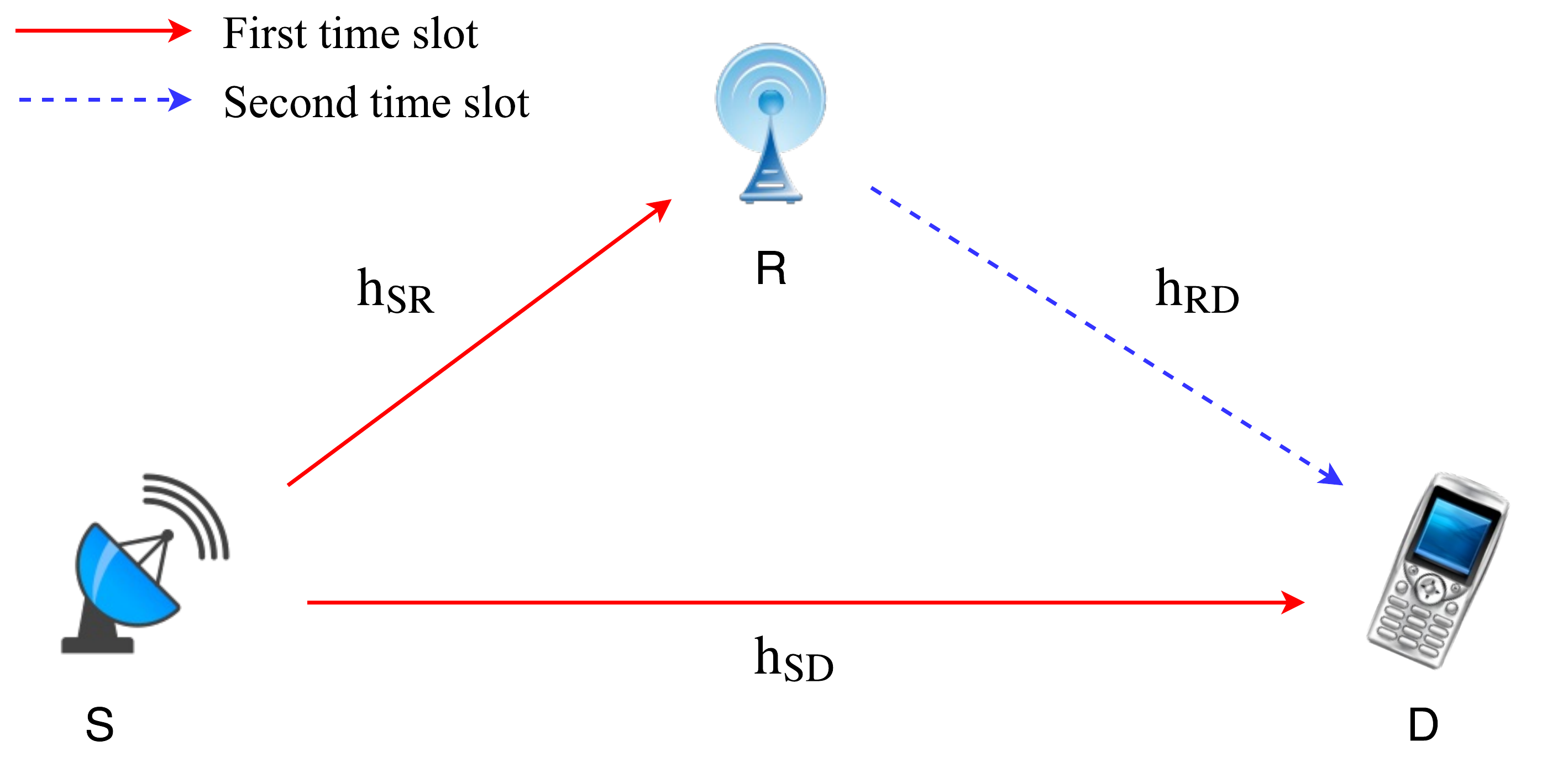}
\caption{Signal transmission with the help of NOMA based cooperative relaying systems when $\lambda_{SD}<\lambda_{SR}$.}
\label{fig1}
\end{figure}
\begin{figure}
\includegraphics[width=\linewidth]{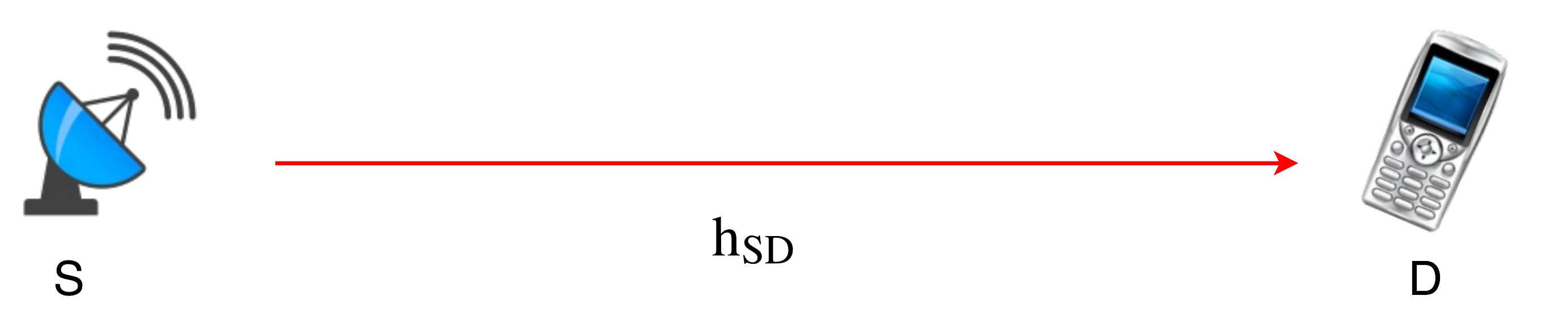}
\caption{Direct signal transmission from the source to the destination when CSI is known at source end i.e. $\lambda_{SD}>\lambda_{SR}$.}
\label{fig2}
\end{figure}
\section{System Description}
\label{sec2}
A CRS has been considered for the analysis as shown in Fig. \ref{fig1}, where a source (S) sends signals to an intended destination (D)
directly and through a half-duplex relay (R). The channel coefficients of S-D, S-R, and R-D links are represented as $h_{SD}$, $h_{SR}$, and $h_{RD}$, respectively, and are considered as independent Rician random variables
with average channel powers represented by $\Omega^2_{SD}$, $\Omega^2_{SR}$ and $\Omega^2_{RD}$, respectively. It is also presumed that $\Omega^2_{SD}<\Omega^2_{SR}$, as usually the path loss of the S-D link, is normally worse than that of the S-R link.

In the CRS using NOMA, in the first time slot, S sends $\sqrt{a_1 P_T}s_1+\sqrt{a_2 P_T}s_2$ to R and D, using the superposition coding \cite{ding2014performance}, where $s_i$ represents the $i$-th symbol, $E[{|s_i|}^2]=1$ and $a_i$ represents the power allocation coefficient for symbol $s_i$ and $a_1+a_2=1$. It is assumed that $a_1 > a_2$, which means
$0.5 < a_1 < 1$ and $0 < a_2 < 0.5$. R decodes $s_2$ after decoding and canceling $s_1$ with SIC, whereas D decodes $s_1$ considering $s_2$ as noise and $P_T$ signifies the total transmit power from S and R both.  In the second
time slot, the only R forwards the decoded $s_2$ with power $P_T$ to D.
However, when $|h_{SD}|^2>|h_{SR}|^2$, the NOMA may not provide a gain of end-to-end achievable rate since the received signal power of symbol  at R is very limited. For O-NOMA, if $|h_{SD}|^2>|h_{SR}|^2$, only direct transmission between S-D is performed without relaying as shown in Fig. \ref{fig2}, otherwise the NOMA is employed with relay \cite{lee2017achievable}. The rationale of the O-NOMA is based on the fact that the achievable rate for the relayed link is the minimum of achievable
rates for the S-R link and the R-D link. That is, when $|h_{SD}|^2>|h_{SR}|^2$, the
direct link achieves better rate performance than the relayed link. In the O-NOMA, hence, the source
directly transmits symbol $s_1$ to the destination with power $P_T$ during a time slot when $|h_{SD}|^2>|h_{SR}|^2$.

In the C-NOMA, the received signals at R during the first time slot is given as 
\begin{equation}
  r_{SR}^C  = h_{SR}(\sqrt{a_1 P_T}s_1+\sqrt{a_2 P_T}s_2) + n_{SR},
  \label{eq1}
\end{equation}
and the received signals at D is
\begin{equation}
  r_{SD}^C=h_{SD}(\sqrt{a_1 P_T}s_1+\sqrt{a_2 P_T}s_2) + n_{SD},
   \label{eq2}
\end{equation}
where $n_{SR}, n_{SD}\sim\mathcal{N}(0,\sigma^2)$ which represents zero mean additive white Gaussian noise (AWGN) with variance $\sigma^2$. The received SNRs of $s_1$, to be decoded for SIC, and $s_2$, to be decoded after SIC, at R are respectively
given from (\ref{eq1}) as
\begin{equation}
  \gamma_{SR,s_1}^C=\frac{a_1 P_T|h_{SR}|^2}{a_2 P_T|h_{SR}|^2+\sigma^2}
   \label{eq3},
\end{equation}
\begin{equation}
  \gamma_{SR,s_2}^C=\frac{a_2 P_T|h_{SR}|^2}{\sigma^2}.
   \label{eq4}
\end{equation}
The received SNR of symbol $s_1$ to be decoded at the destination is given from (\ref{eq2}) as 
\begin{equation}
  \gamma_{SD, s_1}^C=\frac{a_1 P_T|h_{SD}|^2}{a_2 P_T|h_{SD}|^2+\sigma^2}.
   \label{eq5}
\end{equation}
In the second time slot, the received signal at D can be calculated as 
\begin{equation}
  r_{RD}^C=\sqrt{P_T}h_{RD}s_2+n_{RD},
     \label{eq6}
\end{equation}
where $n_{RD}\sim\mathcal{N}(0,\sigma^2)$ and thus the received SNR for $s_2$ is given as 
\begin{equation}
  \gamma_{RD,s_2}^C=\frac{P_T|h_{RD}|^2}{\sigma^2}.
     \label{eq7}
\end{equation}
On the other hand, for direct transmission with $n_{SD}\sim\mathcal{N}(0,\sigma^2)$, the received signal of symbol $s_1$ at the destination and its received SNR are, respectively, given as
\begin{equation}
  r_{SD}^D=\sqrt{P_T}h_{SD}s_1+n_{SD},
  \label{eq8}
\end{equation}
\begin{equation}
  \gamma_{SD,s_1}^D=\frac{P_T|h_{SD}|^2}{\sigma^2}.
  \label{eq9}
\end{equation}
\section{Achievable Rate Calculations}
\label{sec3}
Following the fact that the end-to-end achievable rate of DF relay is dominated by the weakest link \cite{danae1999technical}, the achievable rate of the O-NOMA can be calculated as follows \cite{kim2015capacity}:
If $\lambda_{SD}<\lambda_{SR}$
\begin{equation}
\begin{split}
C^{Pro} & =\frac{1}{2}\textrm{min}\{\textrm{log}_2(1+\gamma_{SD,s_1}^C),\textrm{log}_2(1+\gamma_{SR,s_1}^C)\}\\
& + \frac{1}{2}\textrm{min}\{\textrm{log}_2(1+\gamma_{SR,s_2}^C),\textrm{log}_2(1+\gamma_{RD,s_2}^C)\},
\end{split}
  \label{eq10}
\end{equation}
which can be modified as CSI is available at the source end as \cite{lee2017achievable}
\begin{equation}
\begin{split}
C^{Pro} & =\frac{1}{2}\textrm{min}\{\textrm{log}_2(1+\gamma_{SD, s_1}^C)\}\\
& + \frac{1}{2}\textrm{min}\{\textrm{log}_2(1+\gamma_{SR, s_2}^C),\textrm{log}_2(1+\gamma_{RD,s_2}^C)\},
\end{split}
  \label{eq11}
\end{equation}
and for $\lambda_{SD}>\lambda_{SR}$
\begin{equation}
C^{Pro}= \textrm{log}_2(1+\gamma_{SD,s_1}^D).
  \label{eq12}
\end{equation}

In (\ref{eq10}), the first and the second parts represent the achievable average rates of $s_1$ and $s_2$, respectively. In the first part of (\ref{eq10}), $\textrm{log}_2(1+\gamma_{SR,s_1}^C)$ is needed to consider that R decodes $s_1$ successfully using SIC, but in (\ref{eq11}), $\textrm{log}_2(1+\gamma_{SR,s_1}^C)$ is removed because of the channel condition assumed as $\lambda_{SD}<\lambda_{SR}$. There is one half spectral efficiency penalty for relaying in (\ref{eq11}) but no spectral efficiency penalty is carried for the direct transmission in (\ref{eq12}) as S transmits an independent data symbol directly to D for a given time slot, when  $\lambda_{SD}>\lambda_{SR}$.

Let $\lambda_{SD}\triangleq|h_{SD}|^2$, $\lambda_{SR}\triangleq|h_{SR}|^2$, $\lambda_{RD}\triangleq|h_{RD}|^2$, $\rho=\frac{P_T}{\sigma^2}$ and $C(x)\triangleq \textrm{log}_2(1+x)$, where $\rho$
is the transmit SNR. Hence, using (\ref{eq11}) and (\ref{eq12}), the achievable average rate of the O-NOMA can be obtained for $\lambda_{SD}<\lambda_{SR}$ as
\begin{equation}
\begin{split}
C^{Pro} & = \frac{1}{2}\textrm{log}_2(1+{\lambda_{SD}\rho)}-\frac{1}{2}\textrm{log}_2(1+a_2 \lambda_{SD} \rho)\\
& + \frac{1}{2}\textrm{log}_2(1+\textrm{min}\{a_2\lambda_{SR},\lambda_{RD}\} \rho),
\end{split}  \label{eq13}
\end{equation}
and for $\lambda_{SD}>\lambda_{SR}$
\begin{equation}
C^{pro}= \textrm{log}_2(1+\lambda_{SD}\rho).
  \label{eq14}
\end{equation}
Now, using (\ref{eq13}) and (\ref{eq14}), the total achievable average sum rate of O-NOMA can be calculated as
\begin{equation}
\begin{split}
C^{\overline{Pro}} & =\underbrace{\frac{3}{2}\textrm{log}_2(1+{\lambda_{SD}\rho)}-\frac{1}{2}\textrm{log}_2(1+a_2 \lambda_{SD} \rho)}_{C_{C,{s_1}}}\\
& + \underbrace{\frac{1}{2}\textrm{log}_2(1+\textrm{min}\{a_2\lambda_{SR},\lambda_{RD}\} \rho)}_{C_{C,{s_2}}}.
\end{split}\label{eq15}
\end{equation}
Let, $\gamma_1\triangleq \lambda_{SD}$,  $\gamma_2\triangleq \textrm{min}{\{\lambda_{SR},\lambda_{RD}\}}$, the cumulative distribution function (CDF) of $\gamma_1$ and $\gamma_2$ for $C_{C,{s_1}}$ and $C_{C,{s_2}}$ is given as \cite{7983401}
\begin{equation}
\begin{split}
F(\gamma_1) & = 1-A_x A_y\sum_{k=0}^{\infty}\sum_{n=0}^{\infty}\tilde B_x(n)\tilde B_y(k) n!k!e^{-(a_x+a_y)\gamma_1}\\
& \times \sum_{i=0}^{n}\sum_{k=0}^{k} \frac{a_x^i a_y^j}{i!j!}\gamma_1^{i+j},
\end{split}\label{eq16}
\end{equation}
\begin{equation}\begin{split}
F(\gamma_2) & = 1-A_z A_y\sum_{k=0}^{\infty}\sum_{n=0}^{\infty}\tilde B_z(n)\tilde B_y(k) n!k!e^{-(a_z+\frac{a_y}{a_2})\gamma_2} 
\\
& \times\sum_{i=0}^{n}\sum_{k=0}^{n} \frac{a_z^i {(a_y/a_2)}^j}{i!j!}\gamma_2^{i+j},
\end{split}\label{eq17}
\end{equation}
where, $B_x(n)=\frac{K_x^n(1+K_x)^n}{\Omega_x^n(n!)^2}, B_y(k)=\frac{K_y^k(1+K_y)^k}{\Omega_y^k(k!)^2},  a_x=\frac{1+K_x}{\Omega_x},  a_y=\frac{1+K_y}{\Omega_y}, A_x=a_xe^{-K_x}, A_y=a_ye^{-K_y}, \tilde B_x(n)=\frac{B_x(n)}{a_x^{n+1}},
 B_y(k)=\frac{B_y(k)}{a_y^{k+1}}$. Here, $x$, $y$ and $z$ represent the S-D, S-R, the R-D
links, respectively and $K$ represents the Rician factor.

The approximated values for $C_{C,{s_1}}$ and $C_{C,{s_2}}$ is given in \cite{7983401} and with the help of that, \eqref{eq16} and \eqref{eq17} can be approximated for their asymptotic results as
\begin{equation}
C_{C,{s_1}}=\frac{1}{2\textrm{ln}(2)}[3H(\rho)-H(\rho a_2)],\label{eq28}
\end{equation}
and
\begin{equation}C_{C,{s_2}}=\frac{1}{2\textrm{ln}(2)}G(\rho).
\label{eq31}
\end{equation}
So, the final achievable average rate can be calculated as
\begin{equation}
C^{\overline{Pro}}  = \frac{1}{2\textrm{ln}(2)}[3H(\rho)-H(\rho a_2)+G(\rho)].
\end{equation}
where
\begin{equation}
\begin{split}
H(\rho) & = A_x A_y\sum_{k=0}^{\infty}\sum_{n=0}^{\infty}\tilde B_x(n)\tilde B_y(k) n!k!
\sum_{i=0}^{n}\sum_{j=0}^{k} \frac{(i+j)!}{i!j!}\frac{a_x^i a_y^j}{\rho^{i+j}}\\
& \times e^\frac{a_x+a_y}{\rho}\bigg(\frac{1}{2\frac{a_x+a_y}{\rho}}\bigg)^{i+j}\frac{\pi}{n}\sum_{t=1}^{n}(\textrm{cos}(\frac{2t-1}{2n}\pi)+1)^{i+j-1}
\\
& \times e^{-\frac{2\frac{a_x+a_y}{\rho}}{\textrm{cos}(\frac{2t-1}{2n}\pi)+1}}|\textrm{sin}(\frac{2t-1}{2n}\pi)|,
\end{split}\label{eq29}
\end{equation}
and
\begin{equation}
\begin{split}
G(\rho) & = A_z A_y\sum_{k=0}^{\infty}\sum_{n=0}^{\infty}\tilde B_z(n)\tilde B_y(k) n!k!\sum_{i=0}^{n}\sum_{j=0}^{k} \frac{(i+j)!}{i!j!}\\
& \times \frac{a_z^i{(a_y/a_2)}^j}{\rho^{i+j}} e^\frac{a_z+a_y/a_2}{\rho}\bigg(\frac{1}{2\frac{a_z+a_y/a_2}{\rho}}\bigg)^{i+j}\frac{\pi}{n}
\\
& \times \sum_{t=1}^{n}(\textrm{cos}(\frac{2t-1}{2n}\pi)+1)^{i+j-1}
e^{-\frac{2\frac{a_z+a_y/a_2}{\rho}}{\textrm{cos}(\frac{2t-1}{2n}\pi)+1}}\\
& \times |\textrm{sin}(\frac{2t-1}{2n}\pi)|.
\end{split}\label{eq32}
\end{equation}

\section{Numerical Results and Discussions}
\label{sec4}
\begin{figure}
\includegraphics[width=\linewidth]{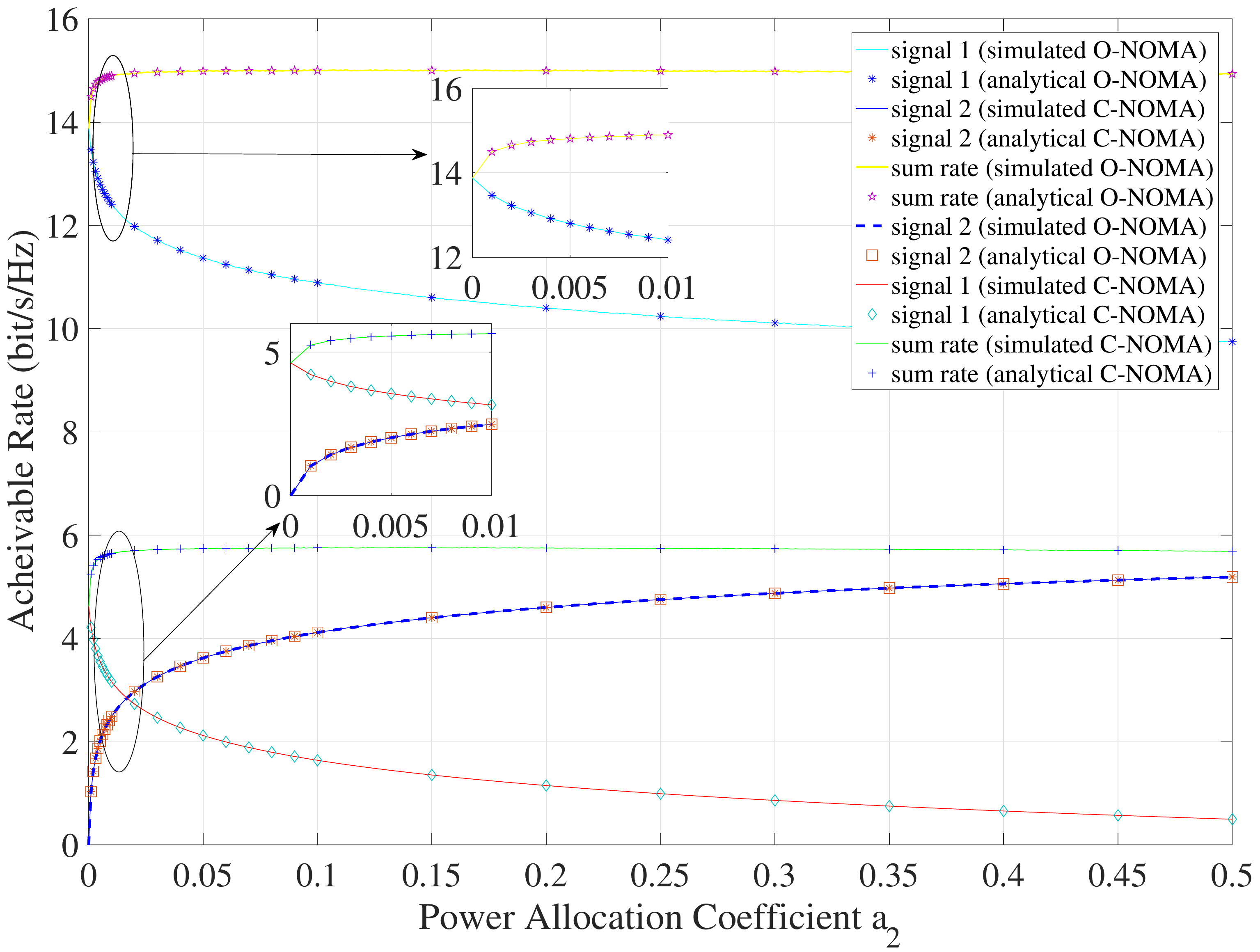}
\caption{Achievable average rates of O-NOMA and C-NOMA with $\rho=20$ dB and $\Omega_{SD}=3$ where $\Omega_{SR}=\Omega_{RD}=6$.}
\label{fig4}
\end{figure}

\begin{figure}
\includegraphics[width=\linewidth]{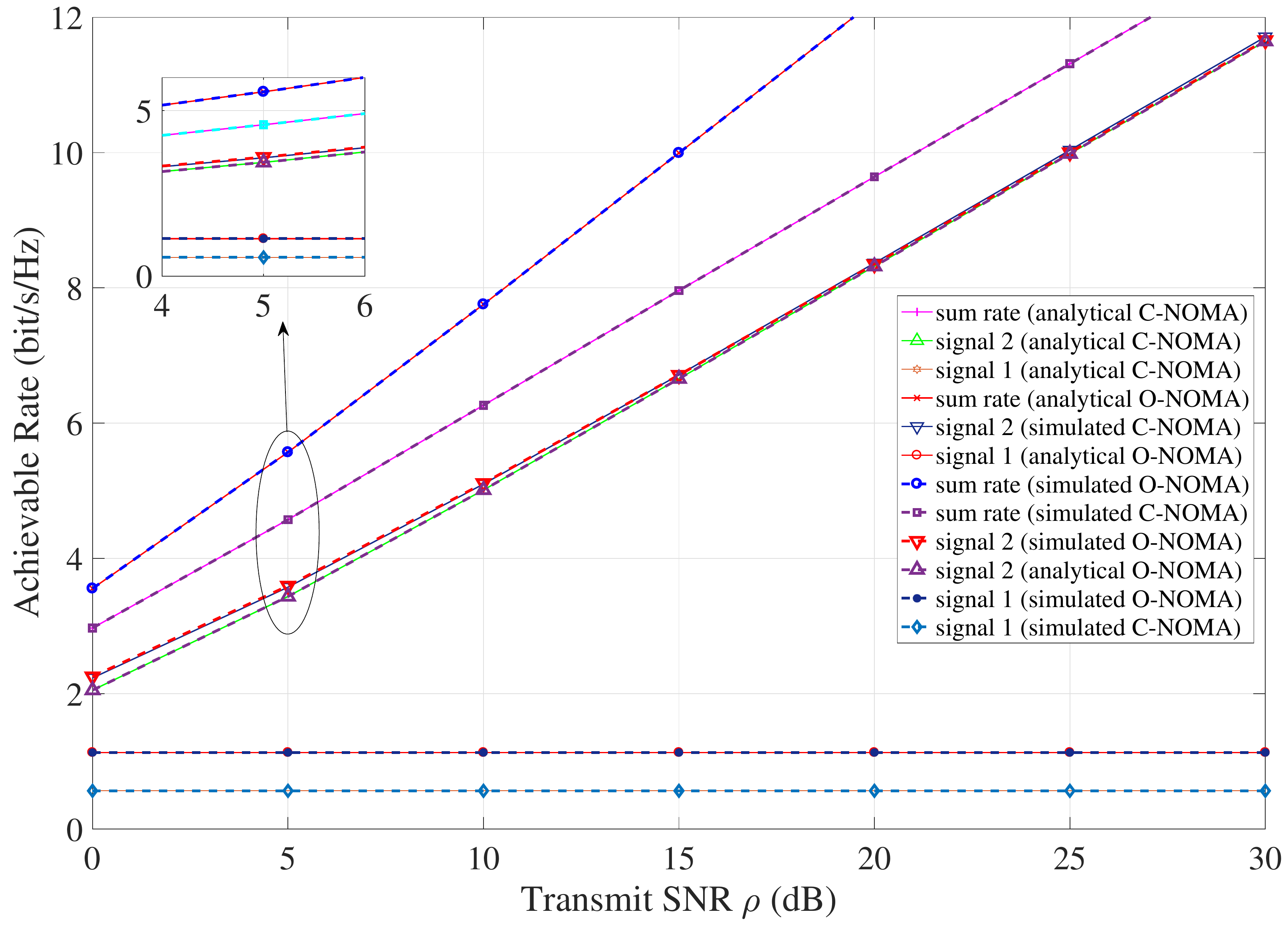}
\caption{Achievable average rates of O-NOMA and C-NOMA for $a_1=0.9$ and $a_2=0.1$ when $\Omega_{SD}=3$ and $\Omega_{SR}=\Omega_{RD}=6$.}
\label{fig8}
\end{figure}

This section presents numerical results to be verified with the results obtained from Monte Carlo simulations.
Fig. \ref{fig4} presents the achievable average rate analysis of $s_1$, $s_2$ and the equivalent achievable sum rate of the O-NOMA for different values of power allocation coefficient $a_2$ of $s_2$, whereas Fig. \ref{fig8} provides the same for C-NOMA over a range of transmit SNR $\rho$. The parameters used for simulations are considered as $K_{SR} = K_{RD} = 4$, $K_{SD} = 3$ and $\Omega_{SD} = 3$. 
Here, with increasing $a_2$, $s_2$ gets more power and hence, the achievable rate of $s_2$ rises and the rate for $s_1$ drops.
\begin{figure}
\includegraphics[width=\linewidth]{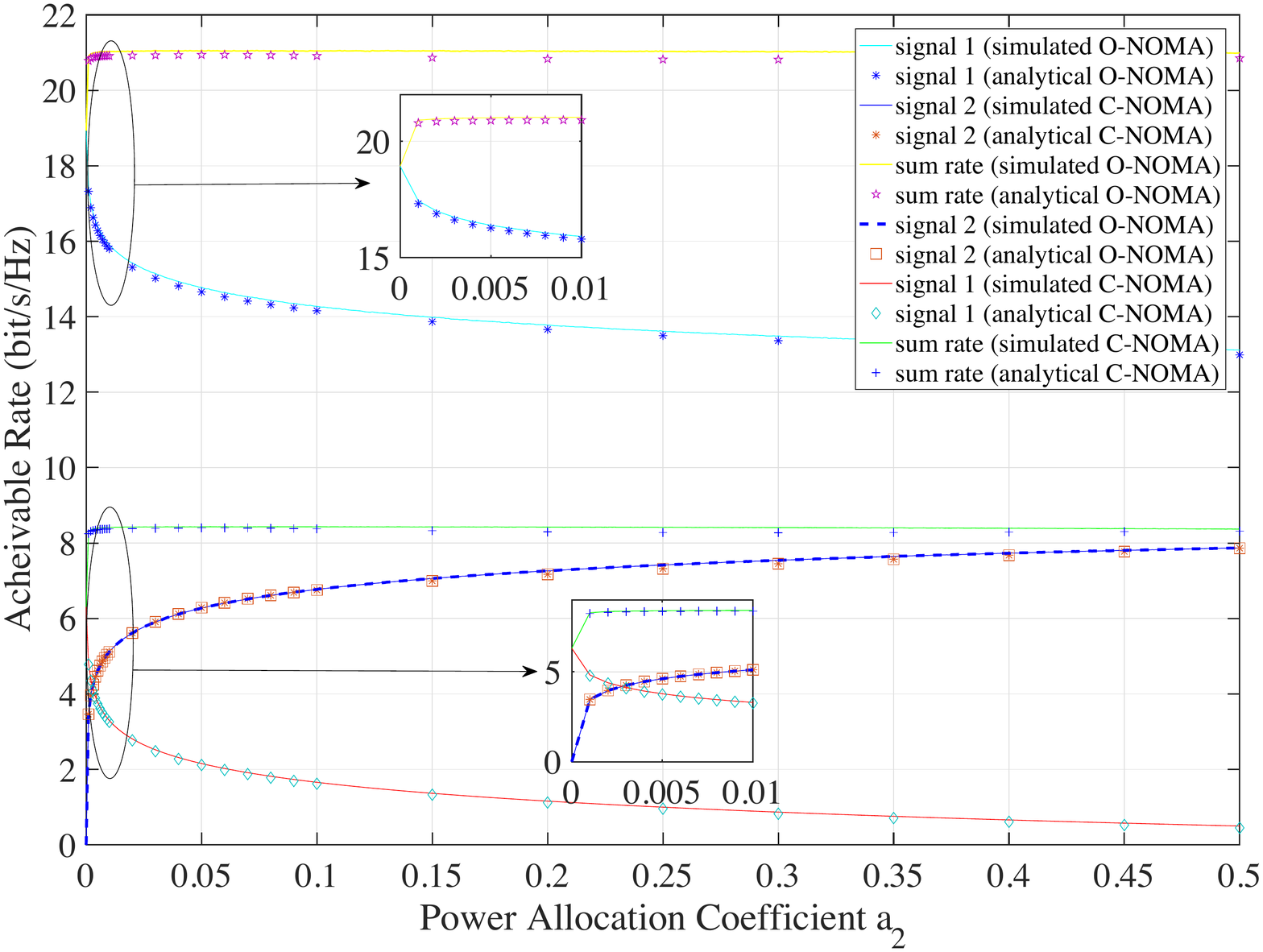}
\caption{Achievable average rates of O-NOMA and C-NOMA with $\rho=30$ dB and $\Omega_{SD}=3$ where $\Omega_{SR}=\Omega_{RD}=12$.}
\label{fig9}
\end{figure}
\begin{figure}
\includegraphics[width=\linewidth]{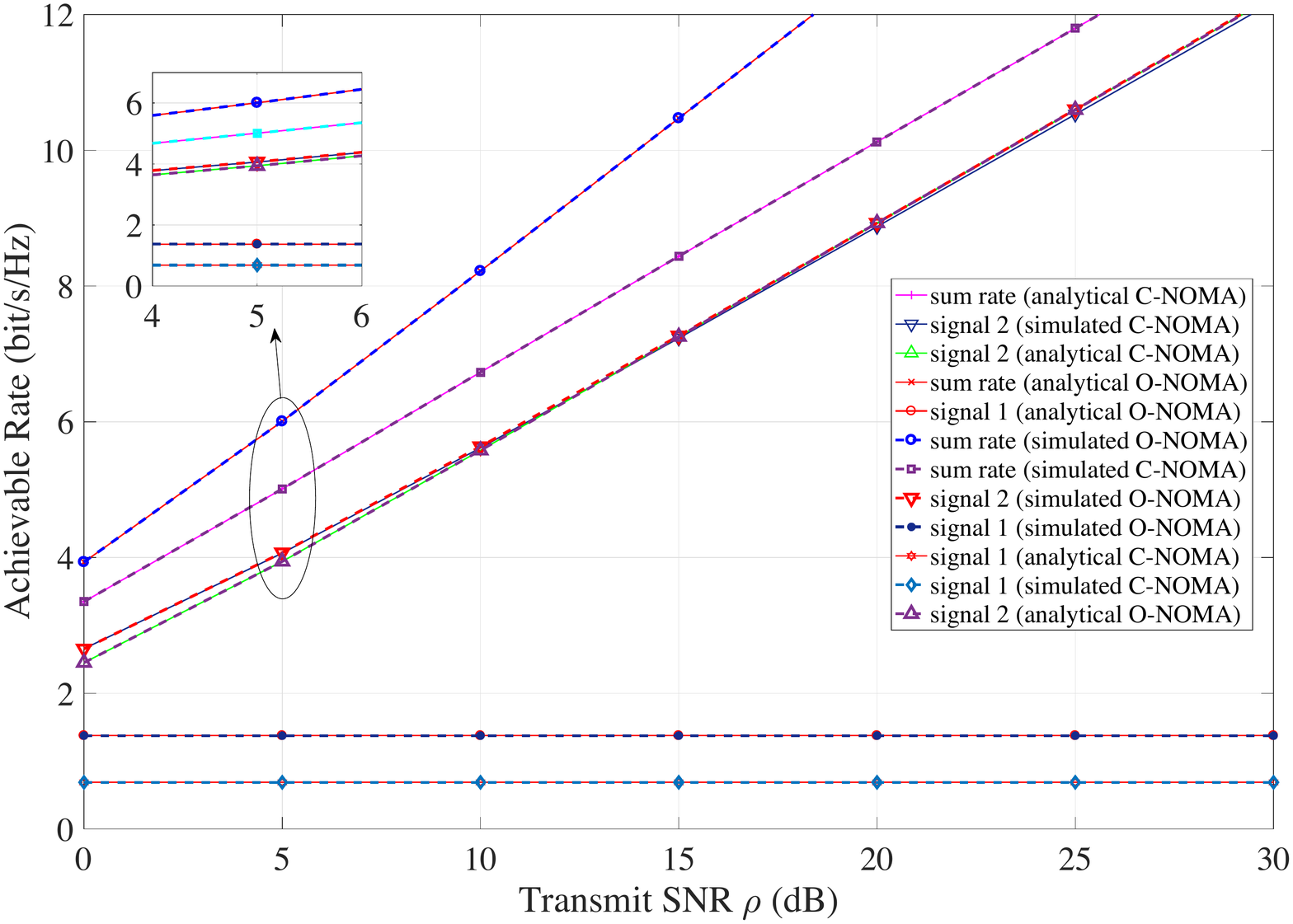}
\caption{Achievable average rates of O-NOMA and C-NOMA for $a_1=0.9$ and $a_2=0.1$ when $\Omega_{SD}=3$ and $\Omega_{SR}=\Omega_{RD}=12$.}
\label{fig10}
\end{figure}

In Fig. \ref{fig4}, $\rho$ is set as $20$ dB and $\Omega_{SR} = \Omega_{RD} = 6$. Now, for O-NOMA, when $a_2$ increases from $0.1$ to $0.4$, the achievable average rate of $s_1$ falls from $10.89$ bit/s/Hz to $9.905$ bit/s/Hz and raises from $4.123$ bit/s/Hz to $5.043$ bit/s/Hz for $s_2$ whereas, for C-NOMA, $s_1$ decreases from $1.64$ bit/s/Hz to $0.657$ bit/s/Hz and the increment in $s_2$ remains the same as in the case of O-NOMA owing to the known CSI at the source. It is clear from the results that $s_1$ achieves $9.25$ bit/s/Hz more achievable average rate in case of O-NOMA in contrast to C-NOMA for $a_1=0.1$. Over and above, achievable sum rates of O-NOMA and C-NOMA are $15$ bit/s/Hz and $5.753$ bit/s/Hz, respectively at $a_2=0.1$, which shows that the O-NOMA achieves $9.247$ bit/s/Hz more achievable average rate than C-NOMA. 

Fig. \ref{fig8} shows that with the increment of $\rho$ from $5$ dB to $15$ dB, the achievable sum rate increases from $5.571$ bit/s/Hz to $9.995$ bit/s/Hz for O-NOMA and from $4.575$ bit/s/Hz to $7.961$ bit/s/Hz for C-NOMA resulting in a gain of $2.034$ for O-NOMA over C-NOMA at $\rho=15$ dB. 
Furthermore, $s_2$ increases from $3.438$ bit/s/Hz to $6.658$ bit/s/Hz for both the systems. At $\rho=15$ dB, it can be seen that $s_1$ is $1.133$ bit/s/Hz for O-NOMA and $0.5663$ bit/s/Hz for C-NOMA. These numerical data verify that O-NOMA is a predominant technique over other existing schemes of CRS-NOMA. 

In Fig. \ref{fig9}
by using $\rho = 30$ dB and $\Omega_{SR} = \Omega_{RD} = 12$, a performance gain of $12.535$ bit/s/Hz over C-NOMA is achieved for $a_2=0.1$, which is $3.288$ bit/s/Hz more than that of the rate achieved for O-NOMA as compared to Fig. \ref{fig4} with $20$ dB of transmit SNR and channel powers of $\Omega_{SR} = \Omega_{RD} = 6$.
Similarly, in Fig. \ref{fig10}, with channel powers considered to be larger as $\Omega_{SR} = \Omega_{RD} = 12$, a gain of $2.03$ bit/s/Hz at $\rho=15$ dB is measured, which is almost equal to the rate gain of O-NOMA over C-NOMA obtained in Fig. \ref{fig8}.

In both the plots, O-NOMA achieves higher rate than the conventional one and indicates that having prior information about the channel state helps to acquire much enhanced spectral efficiency. Furthermore, derived analytical results matched greatly with the results simulated from Monte Carlo simulations, which prove the consistency of the results. Simultaneously, it has also been shown that with the larger channel power of R-D link and increasing SNR, system performance can be improved further.

\section{Conclusions}
\label{sec5}
In this paper, the achievable average rate analysis of Opportunistic-NOMA with Channel State Information at the transmitter node is studied under Rician fading channels. In order to satisfy the goal of the work, a  mathematical expression is also derived from it for its average achievable rate which is plotted against the power allocation coefficient $a_2$ and transmit SNR $\rho$ for its validation. The results are compared with the conventional cooperative relaying scheme C-NOMA with no CSI available at the source terminal. In comparison, derived analytical results show that O-NOMA accomplishes higher performance gain than C-NOMA. Furthermore, the achievable average rate performance of the analyzed O-NOMA is more advanced with the increasing power allocation coefficient $a_2$ of signal $s_2$, transmit SNR $\rho$ and average channel powers. Hence, O-NOMA can be more advantageous and superior to C-NOMA for better spectral efficiency in terms of achievable average rate but requires CSI feedback or any alternative feedback techniques with lower overhead and complexity.

\bibliographystyle{IEEEtran}
\bibliography{references}

\end{document}